\documentclass[a4paper, 12pt]{article}
\usepackage[dvips]{graphics}

\begin{document}

\title{\huge  Phase Transition Critical Flavor Number of QCD.}

\author{ F.  N.  Ndili\\
 Physics Department, University of Houston, \\
Houston, TX.77204-5506, USA.}

\date{August, 2005}

\maketitle

\begin{abstract}
We present an entirely perturbative QCD determination of the
 critical phase transition flavor number $N^{cr}_{f}$ of QCD.
 The results obtained are compared with various determinations of
 $N^{cr}_{f}$ by non-pertrubative methods, including lattice QCD.
  The wider physics implication of the existence of the Banks-Zaks
  regime of QCD with only weakly interacting quarks, is discussed
  briefly.

\end{abstract}

E-mail: ndili@uh.edu

\newpage

\section{INTRODUCTION}
Gies and Jaeckel  in a recent paper [1] raised the question of
reliable \\determination of  the critical flavor number
$N^{cr}_{f}$ separating perturbative and non-perturbative  regimes
of QCD from the high flavor end.  They analyzed the chiral phase
structure of QCD using a functional renormalization group method,
and obtained a value of $N^{cr}_{f} = 10.0 \pm 0.4$.  A number of
earlier determinations of the same quantity by other authors exist
[2-6], based on similar considerations of the chiral phase
transition of QCD.   Non-perturbative  lattice QCD methods have
also been used [7] to find $N^{cr}_{f}$. The importance of
$N^{cr}_{f}$ as a critical phase transition parameter of QCD has
generally been emphasized in these determinations.  By noting that
the take off point common to all the above formulations and
determinations of $N^{cr}_{f}$ is the Banks-Zaks [8] observation
of a weakly coupled two loop infrared fixed point at high flavor
number in QCD, and that this feature of QCD can  be exploited not
only non-perturbatively but also perturbatively, we arrive at the
possibility that $N^{cr}_{f}$ can also be determined by entirely
perturbative methods. Such purely perturbative analysis of the
Banks-Zaks (BZ) infrared phase of QCD were already considered in
another context of searching for the existence of infrared fixed
points in higher order  QCD beta function [9-11]. Our focus in
this paper is on the original two-loop Banks-Zaks infrared fixed
point, presented  as a tool for perturbative determination of the
phase transition critical flavor number $N^{cr}_{f}$ of QCD. The
results so obtained can be compared with various determinations of
$N^{cr}_{f}$ by non-perturbative phase transition methods,
including lattice QCD. With largely consistent firm values
emerging from these different sources,  we discuss the wider
physics implication of the existence of the BZ regime of QCD
and its only weakly interacting quarks.\\

\section{ The Banks-Zaks  Domain of QCD}
A basic observation made by Banks and Zaks in their paper [8] is
that along side the ultraviolet asymptotically free fixed point of
QCD that holds for  $N_{f} \le 33/2$,  there exists close to this
limit point in flavor space, an infrared fixed point that is
induced on QCD by its two loop perturbative beta function. The QCD
beta function is given by:
\begin{equation}\label{eq: ndili2}
\beta (a) = -ba^2 (1 + ca + c_{2}a^2 + c_{3}a^3 + c_{4}a^4 +
\cdots  \rightarrow \infty )
\end{equation}

where $a$ is the QCD coupling  constant in the form $a =
\alpha_{s}/ \pi$.  The first few expansion coefficients $b, c,
c_{2}, c_{3}$ are known, and have the specific values [12-15]:

\begin{equation}\label{eq: ndili3}
b = \frac{33 - 2N_{f}}{6}
\end{equation}

\begin{equation}\label{eq: ndili4}
c = \frac{153 - 19 N_{f}}{2(33 - 2 N_{f})}
\end{equation}

\begin{equation}\label{eq: ndili5}
c_{2}(\overline{MS}) = \frac{3}{16(33 - 2 N_{f})} \left [
\frac{2857}{2} - \frac {5033}{18} N_{f} + \frac{325}{54} N_{f}^2
\right ]
\end{equation}

It is seen that the two loop beta function $\beta (a) = -ba^2 (1 +
ca) = 0$ has an infrared fixed point solution at $ a = a^{*} =
-1/c.$ The fixed point value  depends on flavor, varying from
$a^{*} = 0.012$ at $N_{f} = 16 $ to $a^{*} = 0.21$ at $N_{f} =
8.05. $ For $N_{f} \le 8$, the infrared fixed point vanishes.
Because of the smallness of the QCD coupling constant in this
fixed point domain,  Banks and Zaks  suggested that the domain can
be exploited by making perturbative expansions in the flavor
parameter  $(33/2 - N_{f})$. \\

The BZ  domain has however other features which others have noted.
It is chirally symmetric, and contains massless fermions and
gluons [1-4]. Because of the weak coupling state of the domain,
quarks are not confined there. The domain has no mass gap or
condensate such as would arise from chiral symmetry breaking, and
it has no physical particle spectrum such as form from $q \bar q $
bound states, or Nambu-Goldstone bosons. Finally the domain being
massless  is scale invariant and conformally symmetric [6]. All
these properties provide a variety of avenues by which one can
explore the nature and physics of the BZ domain of QCD, and how
this domain relates  to the rest  QCD. \\

A central point is  that if  QCD  has the  BZ domain controlled as
it is by an infrared fixed point that exists only for certain high
values of flavor number, there must exist in QCD not only a phase
structure, but a critical flavor number  $N^{cr}_{f}$  at which
transitions occur between the different phases of QCD. Banks and
Zaks [8] sketched a phase structure for QCD, but a more authentic
phase structure was later constructed by Appelquist et. al. [2,3].
Based on this QCD phase structure and various non-perturbative
methods, determinations have been made of  $N^{cr}_{f}$ as follows: \\

The value of critical flavor obtained by Appelquist et. al. [2,3]
from their  chiral phase structure and non-perturbative gap
equation for QCD,  is $N^{cr}_{f} = 11.9$. \\

Gies and Jaeckel [1]  combining a model of chiral phase structure
of QCD and Nambu-Jona-Lasinio (NJL) type  four fermion
interactions, find $N^{cr}_{f} =  10.0 \pm 0.4$.\\

Sannino and Schechter [4]  exploring the same chiral phase
structure of QCD, use a non-perturbative effective potential and
perturbative anomalous dimensions, to obtain a value of
$N^{cr}_{f} = 11.7$.  \\

Harada, et. al. [5] from analysis of meson masses in large flavor
Bethe-Salpeter equation obtain a value  for $N^{cr}_{f} = 12$. \\

Miransky and Yamawaki [6] from a general perspective  of conformal
phase transition for non-perturbative dynamics of gauge  fields,
including QCD and its BZ domain, obtain a value for the QCD
critical flavor number  $N^{cr}_{f} = 11.9$. \\

Iwasaki et. al. [7] in a lattice modelling of QCD  chiral phase
 structure  and confinement, obtain a value $N^{cr}_{f} \geq 7$. \\

We  argue that  comparable values for $N^{cr}_{f}$ can be obtained
from an entirely perturbative analysis, based on the original
perturbative expansions suggested by Banks and Zaks in powers of
$(33/2 - N_{f})$. First we note that the methods listed above rely
on a suitable modelling or choice of QCD phase structure or phase
diagram, before one can set up a corresponding non-perturbative
QCD calculation or simulation for $N^{cr}_{f}$. In the
perturbative approach  discussed below, we use only the firmly
established fact that a perturbative infrared fixed point exists
in QCD at two-loop order, and for a range of high flavors.

\section{Purely  Perturbative  Searches for $N^{cr}_{f}$}
The BZ idea is to   exploit the two-loop infrared fixed point near
$N_{f} \le 33/2$  by making perturbative expansions of physical
quantities, around the  point $N_{f} = 33/2$ in flavor space. For
this, we rewrite the two-loop infrared fixed point coupling:
\begin{equation}
a^{*} =  - \frac{1}{c} = \frac{-4}{153 - 19N_{f}} (33/2 - N_{f})
\end{equation}
in the form $ a^{*} =  \epsilon (33/2 - N_{f})$  where $\epsilon =
-4/(153 - 19N_{f}$. We next evaluate $\epsilon$  at $N_{f} =
33/2$, obtaining $\epsilon = 8/321. $ Finally we define a suitable
BZ perturbative expansion parameter  as the quantity:
\begin{equation}
 a_{o} = a^{*}_{min} =    \frac{8}{321} (33/2 - N_{f})
\end{equation}
This quantity $a_{o}$ can have a range of values from $a_{o} = 0$
at $N_{f} = 33/2$ to $a_{o} = 0.21$ at $N_{f} = 8.05$ where the
two-loop infrared fixed point vanishes.  In terms of $a_{o}$, the
two-loop infrared fixed point eqn(6)  becomes reparametrized
 at any one value of $a_{o}$, by the power series:
\begin{equation}
a^{*} =  - \frac{1}{c} = \frac{a_{o}}{1 - (19/4)a_{o}} = a_{o} +
ua_{o}^2 + u^2a_{o}^3 + u^3a_{o}^4 +  .....
\end{equation}
where u = 19/4. Correspondingly :
\begin{equation}
(a^{*})^2 =   a_{o}^2 + 2ua_{o}^3 + 3u^2a_{o}^4 + 4u^3a_{o}^5 +
5u^4a_{o}^6 + 6u^5a_{o}^7 + 7u^6a_{o}^8 + 8u^7a_{o}^9 + 9u^8
a_{o}^{10} + ......
\end{equation}
and
\begin{equation}
(a^{*})^3 =   a_{o}^3 +  3ua_{o}^4 + 6u^2a_{o}^5 + 10u^3a_{o}^6 +
15u^4a_{o}^7 + 21u^5a_{o}^8 + 28u^6a_{o}^9 + 36u^7a_{o}^{10} +
......
\end{equation}

 Any physical quantity expanded perturbatively  in
powers of $a_{o}$  for  any $a_{o}$ value within the above
infrared fixed point range  in the BZ flavor space, should be a
convergent series. While the point $N_{f} = 8.05$ provides a lower
limit below which no such convergence of a perturbative series is
expected, it can happen that  the perturbative series in $a_{o}$
breaks down even before the point $N_{f} = 8.05$ is reached, due
to onset of QCD phase transition boundaries, or to specific
non-perturbative QCD dynamics. Each point of break down of a
physical perturbation series in $a_{o}$ defines a critical flavor
number $N^{cr}_{f} \geq 8.05$ in this NLO case. Thus by studying
various perturbative QCD series reformulated as expansions in
$a_{o}$, and noting the points of break down of each series, one
can deduce corresponding values of $N^{cr}_{f}$ from equation (6).
We now consider some standard QCD perturbation series, and find
their radius of  convergence  in the BZ flavor space. \\

(i) As our first perturbation series we take the ratio quantity: \\
\begin{equation}
R_{e^+e^-}(Q)  =  \frac{\sigma_{tot}(e^+e^- \rightarrow
hadrons)}{\sigma (e^+e^- \rightarrow \mu^+ \mu^-)}
\end{equation}
Its QCD perturbation expansion is  written:  \\
\begin{equation}
R_{e^+e^-}  =  3 \sum_{i=1}^{N_{f}}Q_{i}^2 (1 + R)
\end{equation}
where \\
\begin{equation}
R = a + r_{2}a^2 + r_{3}a^3 + r_{4}a^4 +
  ....
\end{equation}
with [16-20] \\
$ r_{2} = 1.986  - 0.115N_{f}   $ \\
$r_{3} = -6.637 - 1.200N_{f}  - 0.005 N_{f}^2 - 1.2395(\Sigma Q_{f})^2/(3\Sigma Q_{f}^2 )   $ \\

 We evaluate this QCD perturbative series $R_{e^+e^-} $ in the specific
 BZ region  of high flavors and NLO infrared fixed point. For this we set
  $ a = a^{*}$ in equation (12); also we rewrite
each of the coefficients $r_{i}$ in terms of $a_{o}$  using
equation (6). We obtain
\begin{equation}
r_{1} = 0.0885  + 4.6144 a_{o}
\end{equation}

\begin{equation}
r_{2} = -27.798  +  54.77 a_{o}  - 8.05 a_{o}^2
\end{equation}
 The BZ perturbation series for $R_{e^+e^-}$ is now obtained
  by substituting equations (7-9) and equations (13-14) into equation (12).
  We obtain:
\begin{eqnarray}
R_{e^+e^-}  = R(a^*) &=& a_{o} + 4.84a_{o}^2 + 0.2172a_{o}^3 -
184.35a_{o}^4 - 2131.445a_{o}^5 \nonumber\\
&-& 17870.74a_{o}^6 - 13.01468 \times 10^4a_{o}^7 - 87.33858 \times 10^4a_{o}^8 \nonumber \\
&-& 55.5123 \times 10^5a_{o}^9 - 33.9406 \times 10^6a_{o}^{10}
\cdots
\end{eqnarray}

One can next use our requirement that the perturbation series
equation (15) defined as it is in a fixed point domain, must be  a
convergent series, to place a bound on $a_{o}$ and correspondingly
on $N_{f}$.  Before  discussing equation (15) further, we write
down the analogous BZ expansions for other  QCD physical
perturbation series. \\

(ii) Our second example is the Bjorken Sum rule for polarized
electron  nucleon deep inelastic scattering, whose QCD
perturbation series is [21]
\begin{equation}
R_{Bj}^{e} =  a + k_{1} a^2 + k_{2} a^3 +  .....
\end{equation}
where \\
\begin{equation}
k_{1} = \frac{55}{12} - \frac{1}{3} N_{f} =  -0.917  +
13.375a_{o}
\end{equation}
and \\
\begin{equation}
k_{2} = 41.4399 - 7.6073N_{f} + 0.17747N_{f}^2 = -35.7641 +
70.2495a_{o} + 285.729a_{o}^2.
\end{equation}

By plugging equations  (7-9) and (17-18) into equation (16), we
obtain the BZ expansion for the Bjorken  sum rule :
\begin{eqnarray}
R_{Bj}^{e}  = R^{e}(a^*) &=& a_{o} + 3.833a_{o}^2 - 8.5405a_{o}^3
-267.2234a_{o}^4 - 2533.49a_{o}^5 \nonumber\\
&-& 18,929.7a_{o}^6 - 12.6894 \times 10^4a_{o}^7 - 79.8579 \times 10^4a_{o}^8 \nonumber \\
&-& 48.1868 \times 10^5a_{o}^9 - 28.2121 \times 10^6a_{o}^{10}
\cdots
\end{eqnarray}

(iii) As our third QCD perturbation series, we take the Bjorken
Sum rule for neutrino scattering on nucleons. Its QCD perturbation
series is [22]
\begin{equation}
R_{Bj}^{(\nu)} =  a + f_{1} a^2 + f_{2} a^3 +  .....
\end{equation}
where \\
\begin{equation}
f_{1} = 5.75 - 0.44 N_{f}  =  1.58   + 17.833a_{o}
\end{equation}
and \\
\begin{equation}
f_{2} = 54.2298 - 9.4968N_{f} + 0.2392N_{f}^2 = - 37.34588 +
64.332a_{o} + 385.1117a_{o}^2.
\end{equation}

By plugging equations  (7-9) and (21-22) into equation (20), we
obtain the BZ expansion for the Bjorken  neutrino sum rule :
\begin{eqnarray}
R_{Bj}^{(\nu)}  = R^{(\nu)}(a^*) &=& a_{o} + 6.33a_{o}^2 +
18.057a_{o}^3 -  84.315a_{o}^4 - 1360.3677a_{o}^5 \nonumber\\
&-& 11742.8a_{o}^6 - 8.429 \times 10^4a_{o}^7 - 55.2112 \times 10^4a_{o}^8 \nonumber \\
&-& 34.2104 \times 10^5a_{o}^9 - 20.4055 \times 10^6a_{o}^{10}
\cdots
\end{eqnarray}

 (iv) Our fourth case is  the QCD perturbation correction  to
$\tau$  hadronic decay: \\
\begin{equation}
R_{\tau} = \frac{\Gamma ( \tau^{-} \rightarrow \nu_{\tau} +
hadrons)}{\Gamma (\tau^{-} \rightarrow \nu_{\tau} e^- \bar
\nu_{e})}
\end{equation}
 Its QCD perturbation series is [20,23]:
\begin{equation}
R_{\tau} =  a + \tau_{1} a^2 + \tau_{2} a^3 +  .....
\end{equation}
where \\
\begin{equation}
\tau_{1} = F_{3} - \frac{19}{24} \beta_{1}   =  0.08325    +
15.21494a_{o}
\end{equation}
with $F_{3} = 1.9852 - 0.1153N_{f}$ ; and $ \beta_{1} = (2N_{f} -
33)/6 $. Also we have
\begin{equation}
\tau_{2} =  - 38.44269 + 107.0978a_{o} + 254.2058a_{o}^2.
\end{equation}
By plugging equations  (7-9) and (26-27) into equation (25), we
obtain the BZ expansion for the $\tau $  hadronic decay width :
\begin{eqnarray}
R_{\tau}^{(h)}  = R^{\tau}(a^*) &=& a_{o} +  4.83325a_{o}^2 +
0.1223a_{o}^3 - 183.36a_{o}^4 - 1849.236a_{o}^5 \nonumber\\
&-& 13,927.0a_{o}^6 - 9.29142 \times 10^4a_{o}^7 - 5.7986 \times 10^5a_{o}^8 \nonumber \\
&-& 34.647 \times 10^5a_{o}^9 - 20.087 \times 10^6a_{o}^{10}
\cdots
\end{eqnarray}

(v) The last QCD perturbation series we consider is  QCD
correction to scalar Higgs boson hadronic  decay width described
by the  QCD  perturbation series  [24]:
\begin{equation}
R_{H}(a) = 3[ 1 + h_{1} a  +  h_{2}a^2  +  h_{3} a^3 + .......]
\end{equation}
where $h_{1} = 5.66667$ ;
\begin{equation}
h_{2} = 35.93996 - 1.35865N_{f}  =  13.52235   + 54.5158a_{o}
\end{equation}
and \\
\begin{equation}
h_{3} = 164.139 - 25.771N_{f} + 0.25897N_{f}^2 = - 190.57975 +
691.1551a_{o} + 416.952a_{o}^2.
\end{equation}

By plugging equations  (7-9) and (30-31) into equation (29), we
obtain the BZ expansion for the  hadronic Higgs decay width :
\begin{eqnarray}
R_{H}^{(h)}  = R_{H}(a^*) &=& 3 + 17.0a_{o} +  121.37a_{o}^2 +
360.754a_{o}^3 + 47.7a_{o}^4 - 9.4869 \times 10^3a_{o}^5 \nonumber\\
&-& 9.972 \times 10^4a_{o}^6 - 7.742 \times 10^5a_{o}^7 - 52.975 \times 10^5a_{o}^8 \nonumber \\
&-& 33.7675 \times 10^6a_{o}^9 - 20.449 \times 10^7a_{o}^{10}
\cdots
\end{eqnarray}

\section{Analysis of BZ Perturbation Series of \\ Physical Quantities}
We now look at equations (15), (19), (23), (28)  and (32) and make
our deductions. First we comment on the structure of these BZ
series. We notice that the first few terms of each series have
signs opposite to the rest of the infinite series. Furthermore,
simple ratio test of convergence with these early terms does not
yield a consistent result.  Since we are interested in the overall
convergence of each infinite series, we can drop any finite number
of terms without affecting the convergence property of each
series. Accordingly we shall drop the first four terms of each
series and impose our convergence requirement from the fifth term
upwards. As for ratio test of convergence, we  use the more
sensitive Raabe  ratio test [25]. This  states that for a
convergent infinite series of terms: $..... u_{n}, u_{n + 1}
.....,$ the quantity :

\begin{equation}
n \left ( \frac{u_{n}}{u_{n+1}} - 1 \right )  \geq P  > 1
\end{equation}

where n is chosen to be $n \geq N$ and N is a fixed positive
integer, chosen here to be the fifth term of each series. We can
then choose in succession n = 5, 6, ...10 .... in equation (33),
obtaining values of P, and deducing the constraint values on
$a_{o}$. The corresponding constraint values on $N_{f}$ yielding
$N^{cr}_{f}$ are obtained from equation (6).  These values of
$N^{cr}_{f}$ obtained for the various QCD perturbation series we
considered above, are summarized in Table 1. We find a fairly
consistent set of values for the critical flavor number
$N^{cr}_{f}$. These values compare very well with the results
quoted earlier from different non-perturbative QCD methods.  One
feature of our result is that the value of $N^{cr}_{f}$ obtained
for each physical quantity studied,  decreases gradually with
increasing n along each series. The trend is that for $n
\rightarrow \infty$, the value of $N^{cr}_{f} $  converges to 8.05
like $a^*$ itself in eqn.(7). In the process we find we reproduce
all the results of various non-perturbative calculations quoted
earlier.  Our results do not yield the lower lattice QCD limit of
$N^{cr}_{f} \geq 7,$ reported in [7], nor any values $N^{cr}_{f}
\le 6$ suggested in [9,10]. Other independent investigations using
Pade approximants [26] place the QCD infrared phase transition
boundary well above $N_{f} \geq 6$ in agreement with our result.

\section{Summary and Conclusions}
It is certain from several independent  calculations and sources
that the critical phase transition flavor number of QCD lies well
above $N_{f} \geq 8$, associated with the BZ two-loop infrared
fixed point in QCD. \\

 A remarkable fact
coming out of the existence of NLO BZ infrared fixed point region
in QCD, is that quarks of flavor $N_{f} \geq 8$ if they exist at
all, can only be weakly interacting, unable to form bound states
of mesons or hadrons. Therefore they must remain unconfined and
observable in their naked color and fractional charge. While
various searches for conventional quarks  of fractional charge and
naked color appear so far to yield no positive results [27], the
existence of the BZ region of QCD is sufficiently indicative that
these quarks may still be found in the form of very high flavor
quarks, probably in cosmic rays.\\

Additionally, if we remain in the infrared low energy region of
high flavor QCD, the BZ analysis shows that the inter-quark
coupling $q\bar q$ weakens with increasing flavor variety. Thus
quark pairs at $N^{cr}_{f}$ are more strongly paired than quark
pairs of $N_{f} > N^{cr}_{f}$ meaning :
\begin{equation}
\frac{\partial}{\partial N_{f}} <q \bar q >  =  < 0
\end{equation}.

 If this trend is assumed to hold
continuously across the phase transition boundary at $N_{f} \le
N^{cr}_{f}$, one may infer a regime of ordering of  the
inter-quark couplings: $t\bar t < b \bar b <c \bar c < s \bar s <
d \bar d \simeq  u \bar u $. In turn this will reflect  in
observable flavor dependence of  quark condensates, and on the
ratios of these condensates. Dosch [28] for example obtained that
\begin{equation}
\frac{<s \bar s >}{<u \bar u>} = 0.65  < 1.
\end{equation}
 However not enough data presently exist on the values of $<c\bar c >, <b \bar
 b>,\\  <t\bar t>$ condensates, to check the above  BZ extrapolation. Data on these higher
 flavor quarkonium  states, $q \bar q $ potentials and decay widths will also be helpful.\\

\begin{center}
\begin{tabular}{|l|l|l|l|l|l|}
\hline
BZ series & n = 5 & n = 6 & n = 7 & n = 8 & n= 9  \\
 \hline
 $R_{e^+e^-}$ & $N_{f} \geq 12.51$ & $N_{f} \geq 11.78 $ & $N_{f} \geq 11.27$ & $N_{f} \geq 10.89$ &  $N_{f} \geq 10.59$    \\

 \hline
$R_{Bj}^{e}$    & $N_{f} \geq 12.02$   &  $N_{f} \geq 11.37$  & $N_{f} \geq 10.92$   &  $N_{f} \geq 10.59$  & $N_{f} \geq 10.33$   \\

 \hline
 $R_{Bj}^{\nu}$  &  $N_{f} \geq 12.63$  &  $N_{f} \geq 11.71$   & $N_{f} \geq 11.14$  & $N_{f} \geq 10.74$   & $N_{f} \geq 10.45$    \\

\hline
 $R_{\tau}^{(h)}$  &  $N_{f} \geq 12.06$  &  $N_{f} \geq 11.34$  &  $N_{f} \geq 10.87$  & $N_{f} \geq 10.53$  &  $N_{f} \geq 10.27$       \\

 \hline
$R_{H}^{(h)} $  & $N_{f} \geq 13.3$  &  $N_{f} \geq 12.07$ & $N_{f} \geq 11.37$  & $N_{f} \geq 10.90$  & $N_{f} \geq 10.54$     \\

\hline

\end{tabular}
\end{center}

Table caption: Raabe [25] ratios of convergence equation (33),
calculated for various values of $u_{n}/u_{n+1}$ and for different
QCD physical perturbation series. \\

\section{References}
\begin{enumerate}
\item H. Gies and J. Jaeckel, Chiral Phase structure of QCD with
many flavors, hep-ph/0507171, 14 July 2005.

\item T. Appelquist, A. Ratnweera,  J. Terning,  L.C.R.
Wijewardhana, Phys.  Rev. D58,  105017 (1998).

\item T. Appelquist, J. Terning, L.C.R. Wijewardhana, Phys. Rev.
Lett. 77, 1214, (1996).

\item F. Sannino and J. Schechter, Phys. Rev. D60, 056004, (1999).

\item M. Harada, M. Kurachi, K. Yammawaki. Phys. Rev. D68, 076001
(2003).

\item V. A. Miransky and  K. Yamawaki, Phys. Rev. D55, 5051
(1997). (Erratum : Phys. Rev. D56, 3768 (1997)).

\item Y. Iwasaki, K. Kanaya, S. Sakai, and T. Yoshie, Nucl. Phys.
B (Proc Suppl.) 30, 327 (1993).

\item  T.  Banks and  A.  Zaks, Nucl.Phys. B196, 189 (1982).

\item  P.  M.  Stevenson, Phys.  Lett.  B331, 187 (1994).

\item S. A. Caveny and P. M. Stevenson. The Banks - Zaks Expansion
and "Freezing" in Perturbative QCD. hep-ph/9705319.

\item E. Gardi and M. Karliner, Nucl. Phys. B529, 383 (1998)

\item W.  E.  Caswell, Phys.  Rev.  Lett.  33, 244  (1974)

\item D. R.  T.  Jones, Nucl.  Phys.  B75, 531 (1974);\\
    E.  S.  Egorian and O.  V.  Tarasov, Teor.  Mat.  Fiz.  41, 26 (1979)

\item  O.  V.  Tarasov, A.  A.  Vladimirov and A.  Yu.  Zharkov,
Phys.  Lett.  B93, 429  (1980);\\
 S.  A.  Larin and J.  A.  M.  Vermaseren, Phys.  Let.  B303, 334 (1993).

\item T.  van  Ritbergen, J.  A.  M.  Vermaseren and S. A. Larin, Phys. Lett. B400, 379  (1997); \\
 J.  A.  M.  Vermaseren, S.  A.  Larin and T.  van Ritbergen,  Phys.  Lett. B405, 327  (1997).

\item L. R. Surguladze, M. A. Samuel, Phys. Rev. Lett. 66, 560
(1991).

\item M. A. Samuel, L. R. Surguladze, Phys. Rev. D44, 1602 (1991)

\item W. Celmaster, R. J. Gonsalves, Phys. Rev. Lett. 44, 560
(1979).

\item M. Dine, J. Sapirstein, Phys. Rev. Lett. 43, 668 (1979).

\item S. G. Gorishny, A. L. Kataev, S. A. Larin,  Phys. Lett. B259
144 (1991).

\item S. A. Larin, J. A. M. Vermaseren, Phys. Lett. B259, 345
(1991).

\item S. A. Larin, F. V. Tkachov, J. A. M. Vermaseren, Phys. Rev.
Lett. 66, 862 (1991).

\item E. Braaten, S. Narison, A Pich, Nucl. Phys. B373, 581
(1992).

\item K. G. Chetyrkin, Phys. Lett.  B390,  309 (1997)

\item   G. Arfken,  \emph{Mathematical Methods for Physicists},
 p. 245 Second Edition, Academic Press Inc. New York, 1970.

\item  F. A. Chishtie, V. Elias, V. A. Miransky, and T. G. Steele,
Prog. Theor. Phys. 104,  603 (2000).

\item  Particle Data Group: The Euro. Phys. Jour. C3, 349-351
(1998).

\item H. G. Dosch : Prog. Part. Nucl. Phys. 33, 121 (1994). see
also  C. A. Dominguez, A Ramlakan, K.Schilcher Phys. Lett. B511,
59 (2001); Moussallam, Euro. Phys. Jour. C14, 111 (2000).

\end{enumerate}

\end{document}